\documentclass[openacc]{rstransa}

\usepackage{multirow}
\usepackage[normalem]{ulem}

\begin{document}

\title{Buckling of lipidic ultrasound contrast agents under quasi-static load }


\author{
Georges Chabouh$^{1}$, Benjamin van Elburg$^{2}$,  Michel Versluis$^{2}$, Tim Segers$^{3}$, Catherine Quilliet$^{1}$,  and Gwennou Coupier$^{1}$}

\address{
$^{1}$Universit\'e Grenoble Alpes, CNRS, LIPhy, 38000 Grenoble, France\\
$^{2}$Physics of Fluids Group, Technical Medical (TechMed) Center and MESA+ Institute for Nanotechnology, University of Twente, 7500 AE Enschede, The Netherlands\\
$^{3}$BIOS/Lab-on-a-Chip Group, Max Planck Center Twente for Complex Fluid Dynamics, MESA+ Institute for Nanotechnology,
University of Twente, Enschede, Netherlands}

\subject{mechanics, solid state physics}


\corres{Gwennou Coupier\\
\email{gwennou.coupier@univ-grenoble-alpes.fr}}

\begin{abstract}
Collapse of lipidic ultrasound contrast agents under high-frequency compressive load has been historically interpreted by the vanishing of surface tension. By contrast, buckling of elastic shells is known to occur when costly compressible stress is released through bending. Through quasi-static compression experiments on lipidic  shells, we analyze the buckling events in the framework of classical elastic buckling theory and deduce the mechanical characteristics of these shells. They are then compared to that obtained through acoustic characterization.

\end{abstract}

\begin{fmtext}

\section{Introduction}

Ultrasound contrast agents (UCAs) are encapsulated bubbles of a few micrometer in diameter, that have now been widely used in the clinic for microvascular perfusion imaging of the heart, the liver, kidney and other organs \cite{barr2020, frinking2020}. After being intravenously administered, they act as efficient acoustic scatterers owing to their compressibility. They are typically excited by an ultrasonic pulse with a centre frequency close to their MHz Minnaert frequency, the  eigenfrequency of the uncoated bubble \cite{minnaert1933}.

\end{fmtext}


\maketitle

The first generation of UCAs were such free gas bubbles whose effect on a received acoustic signal was discovered by serendipity during an echography of the aorta by Gramiak and Shah \cite{Gramiak68}. As bubbles tend to dissolve rapidly in blood plasma, a second generation of UCAs was developed, made of air bubbles encapsulated by a thin shell: galactose as in Echovist$^{\mbox{\scriptsize{\textregistered}}}$ (1991) or albumin (a human protein) as in Albunex$^{\mbox{\scriptsize{\textregistered}}}$(1995) or galactose and palmitic acid as in Levovist$^{\mbox{\scriptsize{\textregistered}}}$(1995). A third generation of UCAs includes shelled microbubbles with higher life-time, air being simply replaced by a gas with higher molecular weight, of lower solubility: SF$_6$ as  in SonoVue$^{\mbox{\scriptsize{\textregistered}}}$ (2001), C$_3$F$_8$ as in  Definity$^{\mbox{\scriptsize{\textregistered}}}$(2001) or C$_4$F$_{10}$ as  in Sonazoid$^{\mbox{\scriptsize{\textregistered}}}$(2007). The gas core of all these contrast agents  is encapsulated by phospholipids. 


While commercial UCAs are quite polydisperse in size, narrowing the size distribution of UCAs is a way to match better their frequency bandwidth with the relatively narrow one of ultrasonic devices, thus increasing the sensitivity of the whole detection process \cite{versluis2020}. Detection of the point spread function of single UCA is a crucial step that led to the discovery of the Ultrasound Localisation Microscopy \cite{errico2015,couture2018} which is a super-resolution ultrasound imaging technique implemented both in 2D and 3D \cite{chavignon2021}.  
Recently, monodisperse lipid-coated microbubbles were produced using flow focusing devices \cite{parrales14,segers16} and 
they showed 2-3 orders of magnitude higher acoustic sensitivity than that of polydisperse UCAs, be it in-vitro \cite{segers2018} or in-vivo \cite{helbert2020}. To overcome high pressure drops and potential jamming during long operations, a T-junction microfluidic device and ultrasound were combined to produce microbubble suspensions with a narrow size distribution at a high production rate $\sim$ $\num{e6}$ bubbles/s  \cite{carugo2021,khan2022}.
One step further towards versatility, easy storage and transportation was recently achieved by  Soysal et al., who succeeded in preparing monodisperse shelled microbubbles that can be freeze-dried  then de-frozen without apparent change of size or acoustic behaviour \cite{soysal2021}.

\section{Mechanics of UCAs}

Despite their relative simplicity (a spherical air bubble coated by an {\itshape orthoradially} isotropic material), UCAs exhibit a rich dynamics which depends on their size, shell material, and shape of the applied ultrasound waveform \cite{dejong2009}. When excited close to resonance at a centre frequency $f_0$, shelled microbubbles can oscillate non-linearly, {{\itshape i. e.} they} generate echoes at higher harmonics $(2f_0,3f_0...)$, subharmonics $f_0/2$ \cite{sijl2011}, or ultrahamonics $3f_0/2$ \cite{sojahrood2021}. Dollet et al. also reported non-spherical oscillations at pressure amplitudes much below a rupture threshold \cite{dollet2008}. Using the non-linear nature of UCAs, various contrast-enhanced ultrasound imaging techniques have been developed to enhance the signal coming from UCAs especially in the small capillaries embedded in tissues, like harmonic imaging, pulse inversion \cite{couture2008}, contrast pulse sequence \cite{caskey2011} and amplitude modulation. Recently, the latter was used on long sequences with the acquisition starting  without UCAs, in order to later on shutdown the tissue signal \cite{aissani2022}. 

Linear models are able to predict the resonance frequency of a UCA whose shell mechanical properties, including the shell elasticity and the pressure inside the bubble at equilibrium,  have been characterized \cite{van2007,overvelde2010}. These models can also be used to estimate adequate properties from the experimental measurement of the frequency. One popular model is the Hoff model \cite{hoff00}, which consists in the thin-shell limit of the Church model \cite{church95} where a finite thickness shell is considered. It is assumed to be made of a homogeneous, incompressible and isotropic material described by a Kelvin-Voigt model. Due to the small thickness of the lipidic shells, zero-thickness models were also developed by de Jong et al. \cite{deJong92}.

However, using these models to retrieve the shell mechanical properties generally fails to provide consistent results throughout the various experimental techniques used to characterize UCAs. In particular, an apparent increase of shell visco-elastic constants with the radius of the UCA is generally reported, as summarized in \cite{chabouh2021}. As discussed recently in \cite{chabouh2021,dash2022}, considering compressibility or anisotropy  of the material, or refined laws for the surface tension, may help resolve this issue. In both cases, these papers highlight the impact of the initial state (and particularly the inner pressure) on the final results. This points to the necessity of completing spherical oscillation experiments with supplementary mechanical parameters and tests in order to retrieve, beyond effective quantities, genuine mechanical characteristics of a given UCA.

\section{"Buckling" and/or rupture of lipidic UCAs}

Rupture of UCAs under large acoustic stress  has been reported several times in the literature, for lipidic \cite{bouakaz2005} or polymeric ones \cite{helfield2017}. This can be triggered on purpose while using UCAs as drug carriers that would release their cargo \cite{kooiman2014}. While rupture of the shell is the corner stone in drug delivery systems, it is often an undesired behaviour in imaging processes, which needs to be avoided. Using a high speed camera, Bouakaz et al. observed the break-up of polymer microshells when the mechanical index (MI) of an ultrasonic pulse exceeds a certain threshold \cite{bouakaz2005}. Prior to the rupture, authors found a "transient regime" where the microshells may undergo severe size reduction under overpressure without significant expansion during the low pressure half period of the pressure cycles. The MI is defined as the peak negative pressure divided by the square root of the frequency of the ultrasound wave: in this condition, the break-up of UCAs at high MI may be seen as a phenomenon that occurs at high pressure amplitudes if enough time is provided for the associated strong deformation to occur.

Later, various high-speed optical observations of commercial UCAs under acoustic excitation mentioned the appearance of non spherical deformations upon compression. Chetty at al. showed the existence of several folds \cite{chetty2008}, also observed in \cite{marmottant2011}, while Luan et al. highlighted budding and even shedding of small vesicles upon a given compression threshold \cite{luan2014}.

Slower processes can also lead to non-spherical deformations. Dissolution of the inner gas leads to the apparition of several wrinkles \cite{borden2002}, that may be preceded by budding \cite{pu2006,Kooiman2017}.

\subsection{"Buckling" from the surface tension models viewpoint}

Non-linear oscillations of UCAs, including the so-called "compression-only" regime where the shells spend more time compressed than extended  \cite{bouakaz2005}, but also non-spherical deformations upon high enough load may indeed both be accounted for through a model developed by Marmottant et al. \cite{marmottant2005,marmottant2011,versluis2020}. This model is initially based on quasi-static measurement of the mechanical behavior of flat monolayers of lipids; in these experiments, a decrease of the surface tension  $\gamma$ upon compression was observed, until it quasi-vanishes when the available space for each lipid molecules is of order their cross-section. In other words, tension drops when polar heads completely shield the gas-liquid surface interaction. Upon extension, the surface tension reaches a plateau, which is given by the surface tension of the bare interface ({\itshape i. e.} around 70 mN/m for air-water interfaces). It is therefore assumed a constant surface tension above a given surface $A_{ext}$, a zero surface tension below a surface $A_b$, and a linear variation of the surface tension with shell area $A$  in-between, which is (abusively ?) called the "elastic regime". Later-on, smoother parametrization was proposed to allow for analytical resolution of the dynamics \cite{sijl2010,gummer2021}. Segers et al. recently showed that it was possible to  directly retrieve a $\gamma(A)$ curve resembling the ad-hoc assumptions proposed by Marmottant et al. \cite{marmottant2005}, by analyzing the radial oscillations of a shelled bubble under different external static pressures, in order to scan the different equilibrium areas \cite{segers2018}.

The elastic constant associated with this orthoradial tension is the two-dimensional compression modulus $\chi=A d\gamma/dA$, with $A$ the shell area. Due to the small thickness compared to  the shell radius, the bending energy is considered negligible in this model. In the "elastic regime", the elastic constant $\chi$ is therefore proportional to the shell area ; considering the small extent of the elastic regime, this is often presented in the literature as a "constant modulus" regime.

It is important to emphasize a collateral feature of these surface tension models:  the surface $A_b$ at which the elastic modulus vanishes is also that where the surface tension vanishes. 

The vanishing of the elastic constant is a key ingredient to account for the compression-only behaviour and the non-linear modes observed upon a pulsed excitation \cite{sijl2011a,versluis2020}. In parallel, it is understood that the absence of surface tension allows for the appearance of non-spherical shapes of the interface. In the literature, these shapes are qualified as wrinkled, budded, or buckled, while in the initial work of Marmottant et al. \cite{marmottant2005}, they are more schematized as bulged, while being called buckled. On the other hand, buckling is generally defined as a mechanical instability under non-zero load, which triggers well-defined shifts in the geometrical configurations, in order to minimize free energy or enthalpy. In particular, buckling of hollow elastic shells with constant elastic modulus is a well established mechanism. In front of the rich and unexplained collection of non-spherical, dynamical, deformations of lipidic UCAs \cite{bouakaz2005,chetty2008,luan2014,borden2002,pu2006}, the goal of the present paper is to make one step aside and examine how the shell buckling framework can be used to describe the non-spherical deformations of lipidic shells under quasi-static compression.

In particular, the surface tension models state that the tension is always {\itshape positive}, {\itshape i. e.}, the shell would shrink if this tension was not balanced by a positive pressure difference across the shell, from the inside to the outside (the Laplace pressure). Conversely, the classical theory for buckling of elastic shells shows that a shell whose inner pressure is larger than the external pressure is always stable \cite{knoche11} and that unstable branches are met when the external pressure overcomes the inner pressure, {\itshape i. e.} when the shell is under {\itshape negative} tension. While the lipidic shell is compressed until a point where lipids are closely packed ({\itshape i. e.} when surface tension vanishes according to surface tension models), one can consider that repulsive forces will emerge in order to oppose compression, giving rise to an elastic contribution that would be balanced by the external pressure being larger than the internal pressure. The value taken by the internal pressure, and in particular its initial value, is therefore a crucial determinant in this problem.

Another key difference between surface tension based models and elastic materials models, is that neglecting bending energy in surface tension models does not allow for first-order transitions between two different states, but rather the progressive emergence of non spherical shapes while the surface tension progressively decreases, whose geometrical features are dictated by the perturbing field (e. g. thermal, acoustic or hydrodynamic fluctuations).

\subsection{Buckling of elastic shells}

While many experimental results obtained on UCAs activated dynamically by ultrasound have been interpreted using surface tension models, buckling of elastic shells made of an isotropic material has been widely investigated for decades, as has  recently been discussed, e.g. in \cite{hutchinson2016}. Most formulas from shell theory involve the mechanical properties of three-dimensional (3D) shell materials, but one has to keep in mind that the models underlying all calculations consider the thin shell as an elastic {\itshape surface}. In the following, we will limit the mechanical model adequate to describe UCAs using only surface (2D) elastic parameters. This should avoid misconceptions and/or wrong  estimations that would arise from the use of formulas directly implying 3D parameters,  while the objects whose behaviour is described using these tools are not made of an isotropic material, but of a material whose properties and organisation in the shell confer to this latter orthoradially isotropic behaviour (such as a lipidic layer). In particular, while the Poisson ratio in 3D is constrained to $-1\le \nu\le 1/2$ ($\nu = \frac{1}{2}$ corresponding to an incompressible material), these inequalities become $-1\le\nu\le 1$ in 2D, the upper limit corresponding to so-called "incompressible surfaces" (2D shear modulus $\ll$ 2D compression modulus) that is not valid for thin shells of an isotropic material.

In the following, we will reveal the link between buckling features and 2D mechanical parameters. These are the in-plane compression modulus $\chi_{2D}$, a bending modulus $\kappa$ and a Poisson ratio $\nu$. For a spherical shell composed of a homogeneous, isotropic material of finite thickness $d$, Young modulus $E$ and Poisson ratio $\nu_{3D}$, these 3D parameters can be cast into the 2D ones in the thin shell limit $d\ll R$, through the relationships $\nu=\nu_{3D}$, $\chi_{2D}= E d /(2 (1-\nu))$, and $\kappa= E d^3/ (12 (1-\nu^2))$.

How specific mechanical properties will impact the buckling process of microshells has already been questioned for other types of materials. In order to account for the buckling of protein shells like viruses, Ru introduced both the possibility of a bending modulus not related to homogeneous 3D elastic properties, and of a transverse shear modulus that would be much smaller than the in-plane shear modulus \cite{ru2009}. Noteworthy, the concern of Ru was for effective thicknesses $d_{eff}$ smaller than $d$.\footnote{Ru introduced the thickness $d_0$ such that the bending modulus reads $\kappa=Y_{2D} d_0^3/(12 (1-\nu)d)$, leading to $d_{eff}^2=d_0^2/d$, and found a buckling threshold equal to that given in Eq. \ref{eq:DeltaPc_BidiOnly}, in the absence of surface tension.} As buckling is a localized event, introducing an effective thickness is also a way to account for the surface heterogeneities, where buckling would be preferentially triggered. This possibility has been used successfully by Chitnis et al. to account for the rupture of polymer microshells \cite{chitnis2011}. {A few years later, the same group demonstrated that information on the buckling threshold of these shells under quasi-static loading can be used as a qualitative indicator for their response under ultrasound \cite{koppolu2015}.} \\

Here, in order to take into consideration the importance of surface tension effects for lipidic UCAs, we will reconsider thin shell buckling theory in presence of a surface tension $\gamma$. As a first approach, and considering the number of parameters to be determined, we will consider this surface tension to be constant (but potentially non zero), as considered in the Marmottant model in the vicinity of the buckling threshold. {This surface tension is an effective surface tension accounting for all surface effects around the gas-lipid layer-fluid interfaces, a minimalist approach aiming at discussing easily surface tension effects.} Our model will be confronted with the behavior of lipidic UCAs in the vicinity of their potential buckling threshold under quasi-static load. To that aim, here we consider monodisperse microbubbles with different covering shell composition in order to vary their mechanical properties and submit them to a slowly varying external pressure while monitoring their shape deformation through a microscope.

\section{Buckling of an elastic shell in presence of surface tension}
\label{sec:model}
\subsection{Equilibrium in spherical mode}

In spite of its geometrical simplicity, the problem of the instability of a spherical shell requires to enter into cumbersome calculations (see next Sec. \ref{subsec:instab}). {Before introducing them, we establish the equilibrium conditions in the spherical configuration and in the small deformations limit. }

Let us consider a model for a shell of midradius $R$ with spherical
symmetry, taking into account both shell Hookean elasticity (compression modulus $\chi_{2D}$
and a Poisson's ratio $\nu$), a bending modulus $\kappa$ and a surface
tension $\gamma$ due to the enthalpic cost of contact between incompatible
chemical species such as polar-apolar media. The free energy
of the whole object writes :

\begin{equation}
\mathcal{F}\left(R\right)=\frac{1}{2}\chi_{2D}\left(\frac{\Delta S}{S}\right)^{2}\times4\pi R_{ref}^{2}+\gamma\times4\pi R^{2}+8\pi\kappa, \label{eq:NRJlibre}
\end{equation}

where $R_{ref}$ is the radius at which the compressive Hookean part is zero, and
$\frac{\Delta S}{S}=\frac{R^{2}-R_{ref}^{2}}{R_{ref}^{2}}$ is the
relative area change. $R_{ref}$  must not be confused with the initial radius $R_0$, which is that of the shell immersed in the fluid  under atmospheric pressure $P_{atm}$. The first term simplifies at first order in
$\frac{\delta R}{R}=\frac{R-R_{ref}}{R_{ref}}$ as $8\pi\chi_{2D}\left(R-R_{ref}\right)^{2}$,
hence it is straightforward to see that the effect of surface tension, in this low deformation, spherical mode, is to modify the compression modulus from $\chi_{2D}$ to $\chi_{2D}\left(1+\frac{\gamma}{2\chi_{2D}}\right)$

{From the dependency of $\mathcal{F}$ with the volume loss $\Delta V$}, we can infer the relationship at equilibrium
between the pressure drop across the shell, and geometrical and mechanical
features: 

\begin{equation}
P_{ext}-P_{int}=\frac{\partial\mathcal{F}}{\partial\left(\Delta V\right)} =\frac{2\chi_{2D}}{R}\left(1-\frac{R^2}{R_{ref}^2}\right)-\frac{2\gamma}{R}, \label{eq:equilibrium} \end{equation}

where the first term provides the contribution of the in-plane compression
to the pressure drop, and the second term corresponds to the Laplace pressure.

Experimentally, the external pressure $P_{ext}$ is controlled, but
there is no access to the pressure $P_{int}$ inside the shells. It is related to the quantity of gas initally 
trapped in the shell. As we are considering slow variations of external conditions, we consider also that the gas process is slow enough to
be isothermal, and use $\frac{P_{int}}{P_{ref}}=\left(\frac{R_{ref}}{R}\right)^{3}$, where we have introduced the inner pressure  $P_{ref}$ in the reference state $R_{ref}$.

For small deformations ($\frac{\delta R}{R_{ref}}\ll1$), we obtain the shell radius as a function of external pressure:

\begin{equation}
R=A+B P_{ext},\, \mbox{with} \,\left\{
    \begin{array}{lcl}
        A&= & \left(4P_{ref}+\frac{4\chi_{2D}}{R_{ref}}-\frac{4\gamma}{R_{ref}}\right)/\left(\frac{3 P_{ref}}{R_{ref}}+\frac{4 \chi_{2D}}{R_{ref}^2}-\frac{2\gamma}{R_{ref}^2}\right)\\
        B & =&-1/\left(\frac{3 P_{ref}}{R_{ref}}+\frac{4 \chi_{2D}}{R_{ref}^2}-\frac{2\gamma}{R_{ref}^2}\right)
    \end{array}
\right.\label{eq:Pext-de-R_theo}
\end{equation}


In order to go further in the determination of the unknown parameters $\chi_{2D}$, $\gamma$, $P_{ref}$ and $R_{ref}$, one can use supplementary hints obtained through the observation
of the buckling, which are detailed in the following section.

\subsection{Buckling threshold and post-buckling shape}
\label{subsec:instab}

\subsubsection{Critical buckling pressure}

Over a certain pressure drop $\Delta P_{b}=\left(P_{ext}-P_{int}\right)_{c}$
called ``critical pressure'', a thin shell of an isotropic medium undergoes a non-spherical deformation casting the shell into a bowl shape, potentially with several radial folds (like in a deflated beach ball).

The seminal work of Koiter \cite{koiter} has been recently revisited by Hutchinson \cite{hutchinson2016}, in a context of growing interest for the complex instability mechanisms of hollow shells \cite{hutchinson2017,holmes2020,audoly2020}. In order to handle the delicate calculation of the stability analysis of such shells, a simplifying hypothesis is typically made regarding the amplitude of the deformations. Be it following moderate rotation theory or Donnell–Mushtari–Vlasov (DMV) theory, the stability analysis continues to the establishment of a set of two equations between the stresses and the inward radial displacement  $w$ away from the equilibrium spherical state. As we wish to introduce the contribution of a constant surface tension in the problem, we need to go into the detail of the derivation of these equations, which we do following the enlightening book of Ventsel and Krauthammer \cite{ventsel}. We refer to the equations in this reference by the suffix VK and follow their derivation of the two governing differential equations (leading to equations 17.36 VK in the absence of surface tension, and for a more general geometry) in the framework of DMV theory. 

The first equation in 17.36 VK  (or 3.3 in \cite{hutchinson2016} with another convention for the displacement which points outward) is the equilibrium equation ; it involves the Airy stress function $\Phi$ which can be differentiated (Eq. 17.31 VK) to obtain the in-plane normal forces $N_1$ and $N_2$ and the in-plane shear force $N_{12}$. For a purely elastic material, these forces are related to the in-plane strains through Hookean laws (Eqs. 12.45 VK and 12.48 VK). We rewrite these equations, obtained in the thin-shell limit of a 3D homogeneous material, by introducing  directly the 2D elastic constants and by subtracting a constant surface tension $\gamma$ to the normal forces: \begin{eqnarray}
N_i&=& \frac{2 }{1+ \nu}\,\chi_{2D}\, (\varepsilon_i+ \nu \varepsilon_{3-i}) - \gamma, \qquad i=1,2 \label{eq:stress-strain-lin} \\
N_{12}&=&\frac{1-\nu}{1+\nu}\,\chi_{2D}\, \varepsilon_{12}, \label{eq:stress-strain-shear}\end{eqnarray}

where $\varepsilon_{i}$ are the in-plane  linear strain components and $\varepsilon_{12}$ the in-plane shear strain. The Airy function is now related to these new stresses. The stress couple-curvatures relations (12.46 VK), that involve the bending modulus $\kappa$ (noted $D$ in \cite{ventsel}) remain unchanged. 

The equilibrium equation for the shell submitted to a normal surface load directly involves the in-plane stresses $N_i$  (and not a stress-strain relationship) and a bending contribution (Eq. 17.32 VK) therefore it remains unchanged in our context. This leads to the first equation between the Airy function and $w$, in the context of spherical geometry (first equation of  Eq. 19.127 VK together with 19.130 VK): \begin{equation}  \kappa  \nabla^2\nabla^2 w = \frac{1}{R_{ref}}\nabla^2 \Phi - \frac{\Delta P R_{ref}}{2} \nabla^2 w.  \label{eq:DMVequilibrium}\end{equation} 

The second equation that is needed is the compatibility equation, that arises from the reduction from a 3D to a 2D configuration: as the three in-plane strain components are only defined through two in-plane displacements, they must fulfil a compatibility condition.  This compatibility equation (Eq. 17.35 VK) relates the different strain components. In our context, Eq. \ref{eq:stress-strain-lin} shows that the linear strains $\varepsilon_i$ can now be written as the sum of an  elastic strains and a constant term that is proportional to the surface tension $\gamma$. Examination of Eq. 17.35 VK shows that this latter term does not lead to additional contribution in the equation, which finally leads to the same second equation in  Eq. 17.36 VK (or  in Eq. 19.127 VK for spherical geometry): \begin{equation}
    \nabla^2  \nabla^2 \Phi+ \frac{Y_{2D}}{R_{ref}} \nabla^2 w=0, \label{eq:compatibility}
\end{equation}

with $Y_{2D}=2 (1-\nu) \chi_{2D}$ the 2D Young modulus. Therefore, a stability analysis as done in \cite{hutchinson2016} or \cite{ventsel} leads to the same buckling pressure difference across the shell as in the absence of surface tension:

\begin{equation}
\Delta P_{b}=\frac{4\sqrt{ Y_{2D} \, \kappa }}{R_{ref}^2}. \label{eq:DeltaPc}
\end{equation}

Note that this buckling threshold is not what would be obtained by naively replacing $\chi_{2D}$ by the whole compression modulus in presence of surface tension $\chi_{2D}\left(1+\frac{\gamma}{2\chi_{2D}}\right)$.

\subsubsection{Bending modulus through the typical length $d_{eff}$}

While introducing the buckling threshold as an additional information to infer the shell properties, we have also introduced the bending modulus $\kappa$, leaving our initial under-determination unsolved. However, bending energy not only  contributes to set the buckling threshold, but it also controls the final buckled shape. In this configuration, the dimple is circled by a rim
of radius $r$ and width $\ell$ which concentrates the deformation,
both in bending and compression. Following (with 2D parameters) the calculation done by
Landau and Lifschitz \cite{landau86} in the absence of surface tension, the bending energy scales as $\frac{12\left(1-\nu^{2}\right)\kappa r^{3}}{R^{2}\ell}$
while the stretching energy scales as $\frac{Y_{2D}d^{3}\ell^{3}}{R^{4}}$.
The sum of the two terms is minimum for $\ell=\sqrt{R_{ref} d_{eff}}$,
where $d_{eff}=\sqrt{\frac{6\left(1+\nu\right)\kappa}{\chi_{2D}}}$.

This characteristic distance $d_{eff}$ for elasticity drives the deformation of shells in the buckled configuration.
It is equal to the thickness $d$
for a thin shell of an isotropic material. As in \cite{quemeneur2012} who considered lipid vesicles and showed that $d_{eff} \ne d $, we leave here this possibility open and consider $d_{eff}$ as a parameter that still needs to be determined.

In our case in the presence of surface
tension, $\chi_{2D}$ should be replaced by the effective compression
modulus $\chi_{2D}\left(1+\frac{\gamma}{2\chi_{2D}}\right)$, which
leads to 
\begin{equation}
d_{eff}=\sqrt{\frac{6\left(1+\nu\right)\kappa}{\chi_{2D} (1+\frac{\gamma}{2\chi_{2D}})}}.\label{eq:Defdeff}
\end{equation}

Since it drives elastic deformations, this quantity can be evaluated
experimentally on buckled shells, either by looking at the size of the rim or, more conveniently, by examining the presence of radial folds within the dimple. Numerical simulations presented in \cite{quilliet12} have highlighted an empirical geometrical relationship between the number of folds $N_f$ and the characteristic length: \begin{equation} N_f=0.940 \sqrt{R_{ref}/d_{eff} }.
\end{equation}

Through Eq. \ref{eq:Defdeff}, we can now rewrite the bending modulus $\kappa$, which cannot be determined directly, as a function of $d_{eff}$ (and other unknown parameters), and inject it in the expression for the buckling threshold:

\begin{equation}
    \Delta P_{b}=\frac{4\chi_{2D}}{R_{ref}}\sqrt{\frac{1-\nu}{3 (1+\nu)}}\frac{d_{eff}}{R_{ref}}\sqrt{1+\frac{\gamma}{2\chi_{2D}} } . \label{eq:DeltaPc_BidiOnly}
\end{equation}

Using equations (\ref{eq:DeltaPc_BidiOnly}) and (\ref{eq:equilibrium}) right before buckling  provides:

\begin{equation} 
R_{ref}-R_b=\left(\frac{R_b}{R_{ref}}\right)^2\sqrt{\frac{1-\nu}{3\left(1+\nu\right)}\left(1+\frac{\gamma}{2\chi_{2D}}\right)}d_{eff} +\frac{\gamma R_b}{2 \chi_{2D}},
\end{equation}

where $R_b$ is the radius of the sphere at buckling threshold. Since $R_{ref}-R_b$ happens to be small (of order a few \% of the initial radius), in the following we will not make difference between $R_b$ and $R_{ref}$ in the expression of values at first order for the deformation $(R_{ref}-R_b)/R_{ref}$, hence 
\begin{equation} 
R_{ref}=R_{b}\left(1+\sqrt{\frac{1-\nu}{3\left(1+\nu\right)} \left(1+\frac{\gamma}{2\chi_{2D}}\right)}
\frac{d_{eff}}{R_{ref}}+\frac{\gamma}{2\chi_{2D}}
\right).
\label{eq:Rref}
\end{equation}

\section{Experimental methods}\label{sec:apparatus}

\subsection{Monodisperse microbubble production}

The monodisperse microbubble suspensions were produced in a flow-focusing device described in detail in~\cite{segers2018}.The bubbles were formed at a temperature of 55$^\circ$C to minimize bubble coalescence in the outlet of the flow-focusing device~\cite{segers2019}. The lipid coating material comprised DSPC and DPPE-PEG5000 mixed at a 9:1 molar ratio (Corden Pharma, Liestal, Switzerland) and at a total lipid concentration of 12.5~mg per mL of air saturated Isoton (Beckman Coulter Life Sciences, Indianapolis, IN, USA). The lipid dispersion was prepared exactly as described in~\cite{segers2017}. The freshly formed bubbles were initially formed  enclosing a gas mixture of 15 volume \% C$_4$F$_{10}$ in CO$_2$, but once stabilized after CO$_2$ dissolution, they ended full of nearly pure C$_4$F$_{10}$ gas as detailed in~\cite{segers20}. The gas and liquid flows were controlled as described by van Elburg et al.~\cite{vanElburg2021}.

\subsection{Acoustic characterization of the shell stiffness}

These microshells react to an ultrasound pulse  by oscillating at a characteristic resonance frequency that depends on the stiffness that will be called dynamic stiffness $\chi_{dyn}$ in the following --- in order to keep in mind it may differ from that obtained in a quasi-static situation. In the spirit of surface tension based models, this stiffness is associated with variations of surface tension with shell area. It was obtained by fitting a theoretical linear oscillator model to measured attenuation spectra, exactly as described in~\cite{segers2018}. The size distribution of the bubble suspensions were also input to the fitting procedure and were obtained using a Coulter Counter (Beckman Coulter Life Sciences, Indianapolis, IN, USA). The attenuation spectra were measured by transmitting narrowband 30-cycle ultrasound pulses with frequencies ranging from 0.5 up to 5.0~MHz in steps of 100~kHz through the bubble suspension confined by a sample holder (8~mm acoustic path length). The transmit transducer (Olympus V304, 2.25 MHz, 1.88~inch focal distance, 1~inch aperture) was calibrated using a fibre-optic hydrophone (Precision Acoustics). The transmit pulses were generated by a waveform generator (Tabor 8026) and amplified (vectawave, VBA100-200) to a peak negative pressure amplitude of 5~kPa. The receiving transducer (Olympus V307, 5~MHz, 1.93 inch focal distance, 1~inch aperture) was  connected to a digital oscilloscope (picoscope 5444d) to record the attenuated signals from which the attenuation spectrum $\alpha$ was calculated as follows~\cite{Segers2016}: 
\begin{equation}
\alpha=\frac{20}{d} log_{10} \frac{|V_{ref}| }{|V_{bub}|},
\end{equation}
where d is  the acoustic pathlength and $V_{ref}$ and $V_{bub}$ are the amplitudes of the received signal without and with microshells, respectively, at the transmit frequency $f_T$. Microshells of dynamic stiffness ranging from 0.55 to 4.5 N/m were considered in the following. In a forthcoming paper from the Twente groups it will be disclosed how the shell stiffness is controlled.

\begin{figure}
\centering
\includegraphics[width=0.4\textwidth]{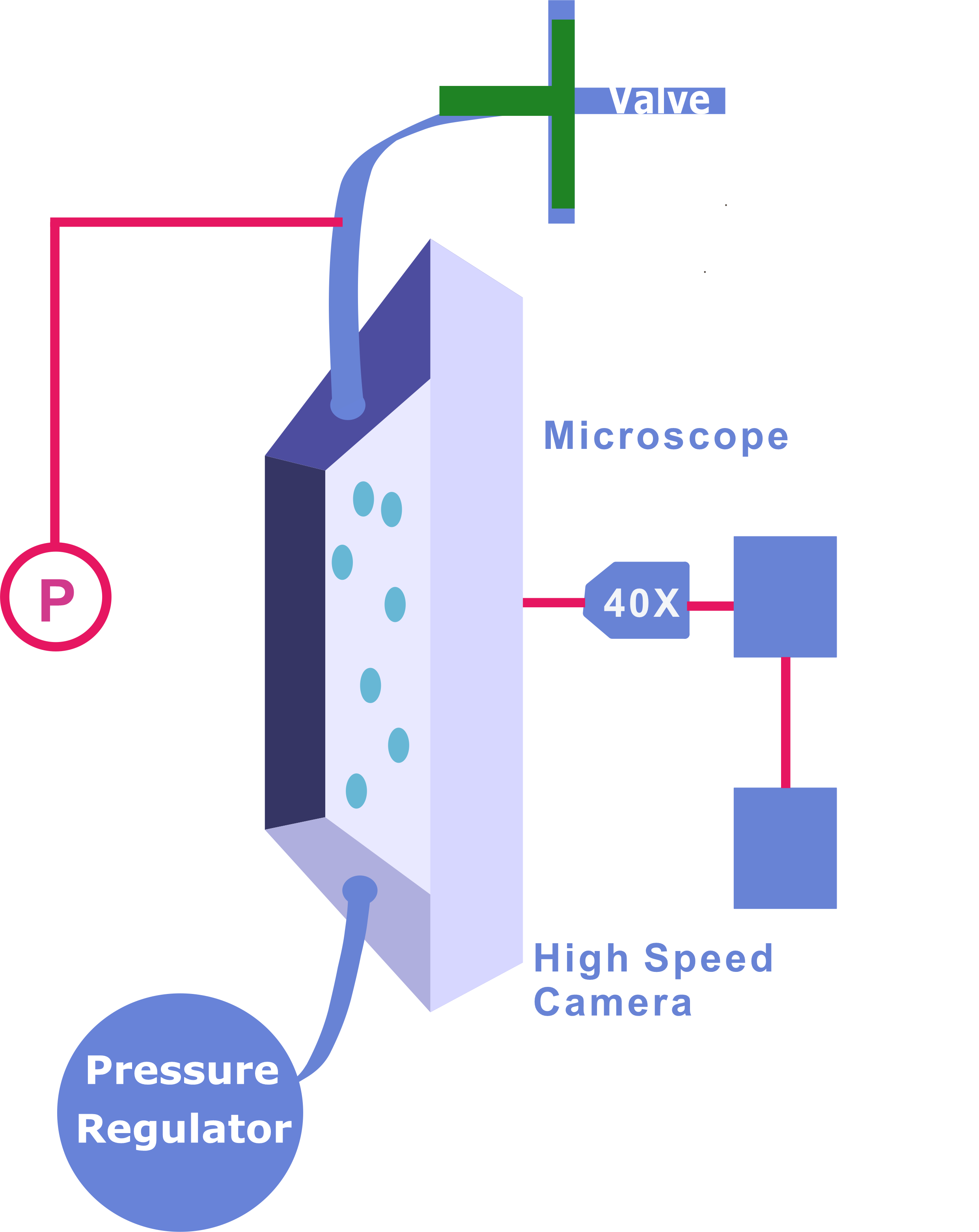} 
\caption{Experimental set-up for buckling experiments. The pressure sensor "P" measures the pressure inside the chamber at the level (according to gravity direction) as the microscope objective. \label{fig:set-up}}
\end{figure}

\subsection{Test case with SonoVue}

In order to check the genericity of our result with our own-made shells, we also used a commercial contrast agent, SonoVue/Lumason$^{\mbox{\scriptsize{\textregistered}}}$ (Bracco Spa, Milan, Italy). It consists in a phospholipid shell encapsulating a sulfur hexafluoride gas core. Different measurements of their visco-elastic properties, obtained through various experimental techniques, can be found in the literature. These parameters are highly dependent on the theoretical model that is used to fit the experimental data, that were also performed at many different conditions (see e.g. effect of driving pressure in \cite{overvelde2010}). The elastic 2D compression modulus is reported to range from $0.024$ to $2.61$ N/m, and the surface viscosity modulus $\kappa_S$ from $\num{0.1e-8}$ to $\num{3e-8}$ N/m.s. 

To maintain good stability of the SonoVue$^{\mbox{\scriptsize{\textregistered}}}$ microshells according to the manufacturer’s suggestions, they were freshly reconstituted prior to use through a mixture of the lyophilisate with 5 ml physiological saline solution, to form a suspension that contains approximately $2-5\times10^{8}$ shells per milliliter with diameters ranging from $2$ to $7$ $\mu$m.

\subsection{Shell pressurization and observation}

Shells were gently poured into degased water and placed in a  Falcon microfluidic reservoir of 15 ml (Fisher Scientific, USA) connected to an Elveflow$^{\mbox{\scriptsize{\textregistered}}}$ pressure controller (Elvesys$^{\mbox{\scriptsize{\textregistered}}}$, France) and to a flow-through cuvette (Aireka Scientific$^{\mbox{\scriptsize{\textregistered}}}$ Co., Ltd) using PTFE tubings. 
The chamber is made from quartz with a square cross section (12.5mm$\times$12.5mm) and it is placed under an inverted microscope (Olympus$^{\mbox{\scriptsize{\textregistered}}}$, model IX70) which was twisted $90^{\circ}$ through three stabilizing aluminium legs. This configuration allows to have an observation axis $z$ perpendicular to gravity axis $y$. After the injection in the chamber, the microshells float up due to buoyancy (see Fig. \ref{fig:set-up}).

The other end of the observation chamber is connected 1) to a valve which is left open to inject the UCAs into the chamber by a gentle increase of the pressure in the reservoir (of order 30 mPa above atmospheric pressure), and closed to allow for pressurization of the chamber in order to compress the shells  and 2) to a pressure sensor (MPS1, Elvesys$^{\mbox{\scriptsize{\textregistered}}}$, France).  Following the instruction of the manufacturer, the sensor was first calibrated by connecting it directly to the outlet of the pressure controller. This sensor was placed at the same altitude as the objective of the microscope and allowed us to measure the ambient pressure around the shells, while checking potential time delay in the response  of the whole device regarding imposed pressure variation.

In order to correlate shell shape evolution with values of external pressure, both pressure sensor and fast camera (Miro 310, Vision Research) are triggered through the Elveflow interface. Movies are taken at a rate of 100 frame/s. An automated stage (MS-2000, ASI, USA) is used to select a microshell prior to the recording.

Experiments were conducted at room temperature (around 25 $^\circ$C) and the concentration of shells ($\sim 10^5$ shells/mL) is such that interactions between shells can be considered negligible.

\begin{figure*}
\begin{tabular}{cc}
      \includegraphics[width=0.4\textwidth]{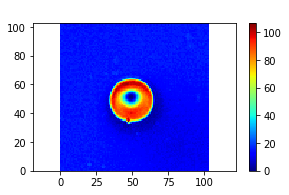} &
      \includegraphics[width=0.4\textwidth]{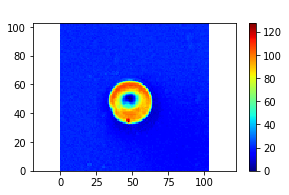} \\
      (a)  & (b)  \\[6pt]
      \includegraphics[width=0.4\textwidth]{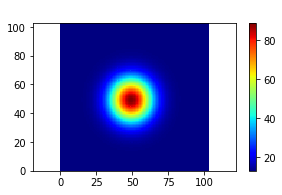} &
      \includegraphics[width=0.4\textwidth]{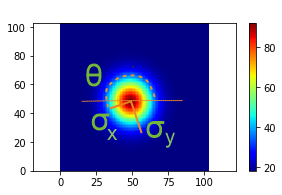} \\
      (c)  & (d)  \\[6pt]
\end{tabular}
\caption{Light intensity distribution of a)  raw image for 'spherical' shape , b) raw image for 'buckled' shape, c) fitted image for 'spherical' shape and d) fitted image for 'buckled' shape. In the spherical phase: $\sigma'_x=5.11$ $\mu$m and $\sigma'_y=5.38$ $\mu$m, in the buckled phase: $\sigma'_x=4.87$ $\mu$m and $\sigma'_y=5.30$ $\mu$m.
 \label{fig:intensity}}
\end{figure*}

\subsection{Image analysis}\label{tracking}

To track the UCAs and get their shape at each time step, we  developed a  tracking algorithm with sub-micron resolution.  The intensity profile of a non-spherical shell is complex and depends on the angle of the shell relative to the optical axis. Being interested essentially in the detection of non-spherical deformation events, we approximate it by a two-dimensional (2D) elliptical Gaussian distribution with a tilt angle $\theta$ \cite{mann1999}. The intensity is expressed as: 

\begin{eqnarray}\label{Eq:3-1}
I(x,y) = I_0 + A \times exp\left( -\frac{x'^2}{2\sigma'^2_x}-\frac{y'^2}{2\sigma'^2_y} \right), \mbox{with}
\end{eqnarray}
\begin{eqnarray}\label{Eq:3-2}
\left\{
\begin{array}{ll}
x' = (x-x_0)cos\theta-(y-y_0)sin\theta\\
y' = (x-x_0)sin\theta+(y-y_0)cos\theta
\end{array}
\right.
\end{eqnarray}
where $(x_0,y_0)$ is the center coordinate of the Gaussian model, $\theta$ is the rotation angle, $\sigma'_x$ and $\sigma'_y$ are the standard deviation of $x'$ and $y'$ axis respectively, $A$ is the Gaussian amplitude peak and $I_0$ is the background amplitude. The fitting parameters are estimated according to the least squares principle. After the fitting, data are processed such as $\theta$ is always the angle of the short axis {\itshape i. e.} min($\sigma'_x$,$\sigma'_y$). $\theta$ is also chosen such that when the ratio $\sigma'_x/\sigma'_y$ is minimal, the corresponding angle $\theta$ indicates the buckling direction, called $\theta_b$ hereafter.
An example of the fitting is shown on Fig.\ref{fig:intensity} where we show the initial raw and fitted image of the same shell in spherical and buckled states respectively.

For comparison with known deformation processes of buckling elastic shells \cite{djellouli2017}, we also defined the (optical) width $w$ and height $h$ of the shell, relative to the buckling angle $\theta_b$ (as defined below), as the widths of the 1D Gaussian profiles in the $\theta_b$ and $\theta_b+\pi/2$ direction, respectively. For a purely axisymmetric, non noisy, process, these values are equal to $\sigma'_x$ and $\sigma'_y$, respectively.

\section{Results}

\begin{figure*}
\begin{center}
       \includegraphics[width=0.5\textwidth]{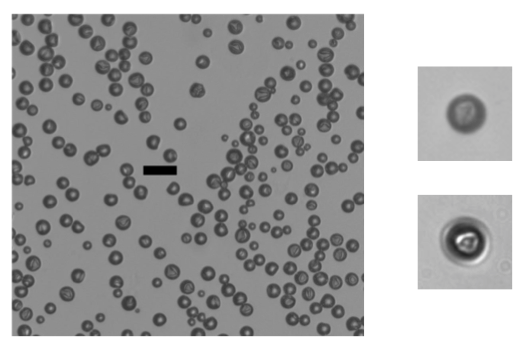}
\end{center}
\caption{Left: Zoology of shapes of home-made UCAs of dynamic modulus  2 N/m after buckling. The 3-fold geometry is dominant. The scale bar represents 6 $\mu$m. Right: buckled shapes  of two Sonovue shells of similar radius, also characterized by a 3-fold geometry.
 \label{fig:shapes}}
\end{figure*}

We first considered slow variations of pressure, which varied from the atmospheric pressure $P_{atm}=101.3$ kPa to a maximum value $P_{\max}$ and back to $P_{atm}$ within 40 s. We systematically observed that when the maximal pressure $P_{\max}$ was set to be slightly above a given value $P_b$, shells would lose their spherical symmetry at threshold $P_b$ and collapse into a bowl-like shape which had most of the time a three-fold symmetry, as can be seen on Fig \ref{fig:shapes}. This deformation was reversible and could be repeated several times through slow pressure cycles between $P_{atm}$ and $P_{max}$ with no apparent damage. {These points are clear signatures of an elastic behaviour, since shapes of shells driven only by surface tension effects would be non-specific when surface tension vanishes}. Threshold  $P_b$ is determined from the rapid change of the apparent aspect ratio $\lambda=\sigma'_x/\sigma'_y$, and will be called  "buckling pressure" in the following, due to similarity between shapes observed on UCAs and typical buckled shapes of elastic shells \cite{knoche11,quilliet12,Knoche2014,coupier2019}.

\subsection{Position of the buckling spot}

A perfectly spherical and homogeneous shell will a-priori buckle {everywhere and simultaneously}. In practice, buckling instability develops around a preferred location corresponding to a "weaker" point, {\itshape i. e.} a thinner or a smoother part of the shell, as demonstrated theoretically or on macroscopic shells designed on purpose \cite{lee16,djellouli2017,yan2020,Stein-Montalvo21,mokbel21}. For lipidic UCAs that do not a priori exhibit well defined defects, the question of the location of the buckling spot is open. In order to answer this question, we observed sequences of buckling events on several shells, that were let free to move far from the chamber walls. Due to buoyancy, they could be observed for around 20 s without moving the microscope stage. In the absence of markers on the shell membrane, we measured for each pressure cycle the angle $\theta_b$ of the buckling direction as defined by the Gaussian shape characterization. In order to keep a precise and reliable measurement of the buckling spot position, we opted for not moving the microscope stage along the process, but rather went for a $f=$2 or 1 Hz pressure cycle in order to obtain around 20 successive measurements of the buckling angle along one trajectory.

\begin{figure*}
\begin{tabular}{cc}
       \includegraphics[width=0.5\textwidth]{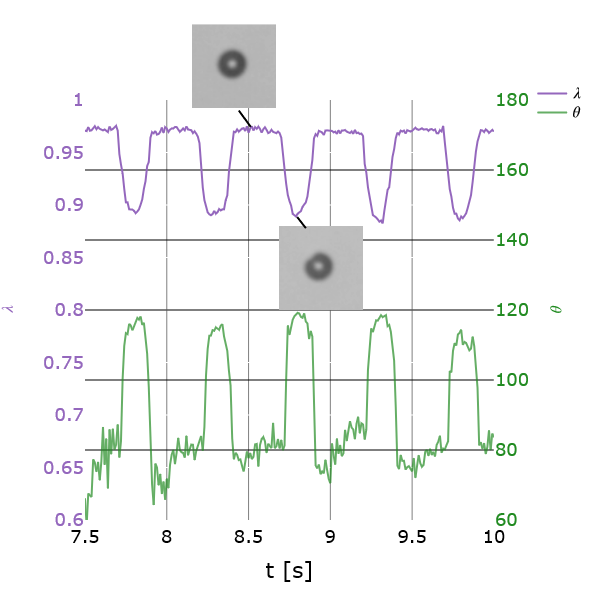} &
       \includegraphics[width=0.5\textwidth]{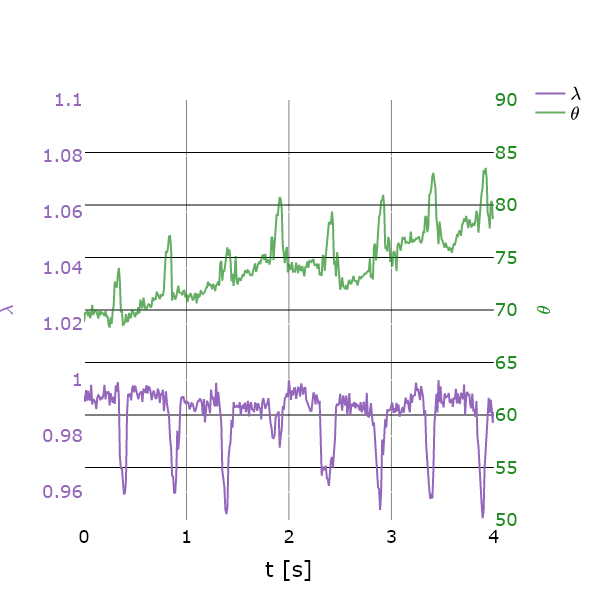} \\
      (a)  & (b)  \\[6pt]
\end{tabular}
\caption{Examples of time-evolution of aspect ratio $\lambda$ (left-axis) and angle $\theta$ (right-axis, in degrees) for two different SonoVue$^{\mbox{\scriptsize{\textregistered}}}$ microshells of initial radius a) 2.5 $\mu$m and b) 2 $\mu$m . External pressure is varied at a frequency of 2 Hz. Pictures in insert show spherical and buckled configuration, right after buckling, when buckling angle $\theta_b$ is determined. 
\label{fig:exampletheta}}
\end{figure*}

In Fig.\ref{fig:exampletheta}, we plot an example of the aspect ratio  $\lambda$  and the fitting angle $\theta$ during pressure cycles. We can clearly see that $\lambda$ varies from a value close to 1 in the spherical shape \footnote{It is slightly smaller than 1 due to non-spherical distribution of the light intensity around the shell.} and suddenly drops. The drop of $\lambda$ corresponds to the buckling, as shown on the inserted snapshot and is thus used to define the buckling angle  $\theta_b$.

\begin{figure*}
\begin{tabular}{cc}
     \includegraphics[width=0.5\textwidth]{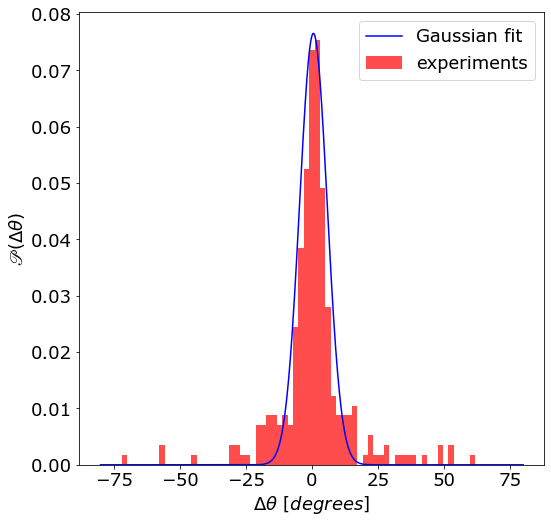} & 
     \includegraphics[width=0.5\textwidth]{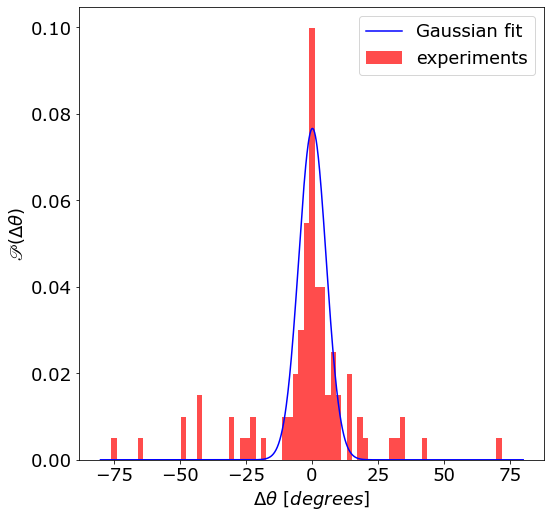}  \\
      (a)  & (b)  \\[6pt]
      \includegraphics[width=0.5\textwidth]{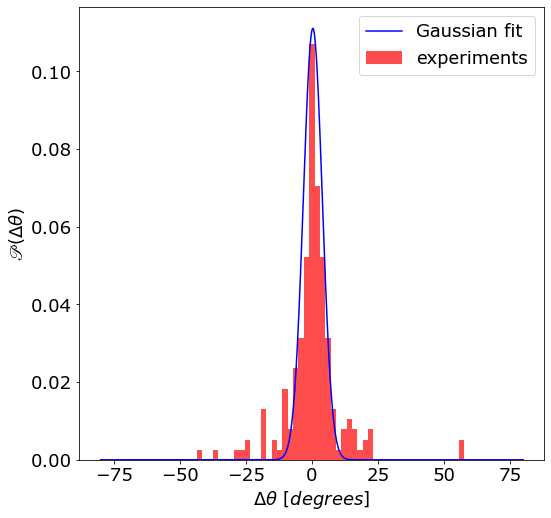} & 
     \includegraphics[width=0.5\textwidth]{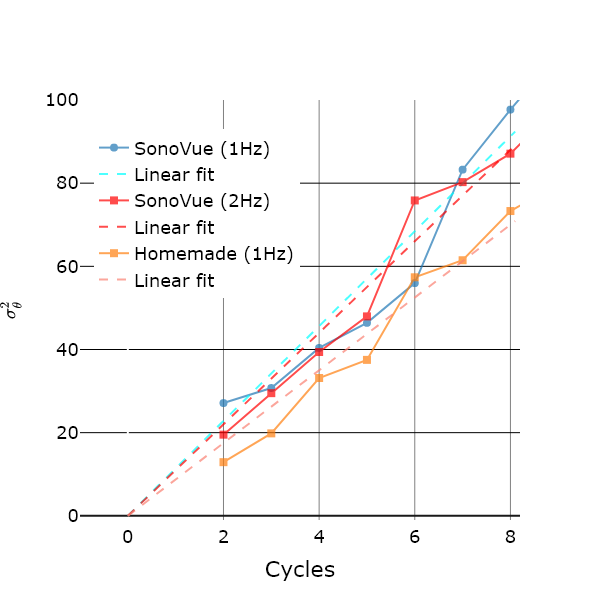}  \\
      (c)  & (d)  \\[6pt]
\end{tabular}

\caption {Probability distribution  $\mathscr{P}(\Delta \theta) $ of the buckling angle variation $\Delta \theta$ after one cycle. (a)  SonoVue ($f=2$Hz, $\bar{R}_0=2.07 \pm0.41$ $\mu$m) ; (b) SonoVue ($f=1$Hz$, \bar{R}_0=2.21 \pm0.38$ $\mu$m); (c) Homemade shell  ($\chi_{dyn}$=2 N/m, $f=1$Hz, $\bar{R}_0=2.07 \pm0.26$ $\mu$m). (d) Squared widths of the Gaussian fits of the distributions of buckling angle variations  after $n$ cycles as a function of the number $n$ of cycles. The corresponding slopes $\tilde{D}$ are $11$, $11.41$ and $8.75$ respectively. }

\label{fig:diffusion}
\end{figure*}

As we cannot track material points on the shell, we check the hypothesis of single buckling location by comparing the time-evolution of the buckling angle with that of a material point on sphere in case of rotational thermal diffusion. The rotational diffusion coefficient $D_r$ of a single colloidal sphere with radius $R_0$ suspended in a solvent with shear viscosity $\mu$ is given by the familiar Stokes–Einstein–Debye relation (\cite{einstein1906,debye1929}):

\begin{eqnarray}\label{Eq:Brownian-rot}
    D_0^r=\frac{k_BT}{8 \pi \mu R_0^3},
\end{eqnarray}
with $k_BT$ the thermal energy. The mean-square angular deviation varies linearly with time and with $2D_0^r$ as a coefficient of proportionality. If there is only one buckling spot, we then have:

\begin{eqnarray}\label{Eq:3-4}
    \langle\theta_b^2\rangle=2D_0^rt
\end{eqnarray}

Considering a narrow size distribution in each case, we show the histograms of $\theta_b$ increment after one cycle on Fig. \ref{fig:diffusion}. While on some rare occasions, the increments can reach quite high values, most are centered around 0. Similar graphs can be obtained after $n$ cycles. Due to the length of the trajectories, we restrict our statistical analysis to $n \le 8$. In what follows, we exclude the data where $\Delta \theta_b$ exceeds $\pm 40$ degrees and we fit the distribution with a Gaussian function. 

The variances $\sigma_\theta^2$ of the Gaussian distributions are plotted as a function of cycle number in Fig. \ref{fig:diffusion}(d), for SonoVue shells excited at 2 or 1 Hz and for home-made shells ($\chi_{dyn}=2$ N/m excited at 1 Hz. They all have a narrow size distribution centered on mean radius $\bar{R}_0$. Variances increase linearly with time, indicating a diffusive behaviour. From the measured slopes $\tilde{D}$, we find that the associated diffusion coefficient should be given by $\bar{R}_0^3 D_0^r=\bar{R}_0^3 \frac{\tilde{D}}{2}\times(\frac{\pi}{90})^2 f= 1.2\times 10^{-19}, 0.8\times 10^{-19}$ and $0.5\times 10^{-19}$ m$^3$.s$^{-1}$, respectively. These values must be compared with  $k_B T/(8\pi \mu)\simeq 1.6\times 10^{-19}$ m$^3$.s$^{-1}$, where $k_B T \simeq 4.1\times 10^{-21}$ J and $\mu=1$ mPa.s. The proximity between these values support the hypothesis  that shells mostly have a unique buckling spot. Whether this unique spot is due to an intrinsic defect, or to a dynamical process where not all stresses have time to relax between two pressure increases, remains to be discussed.

\subsection{Buckling upon a quasi-static loading}

\begin{figure*}
\begin{tabular}{cc}
      \includegraphics[width=0.5\textwidth]{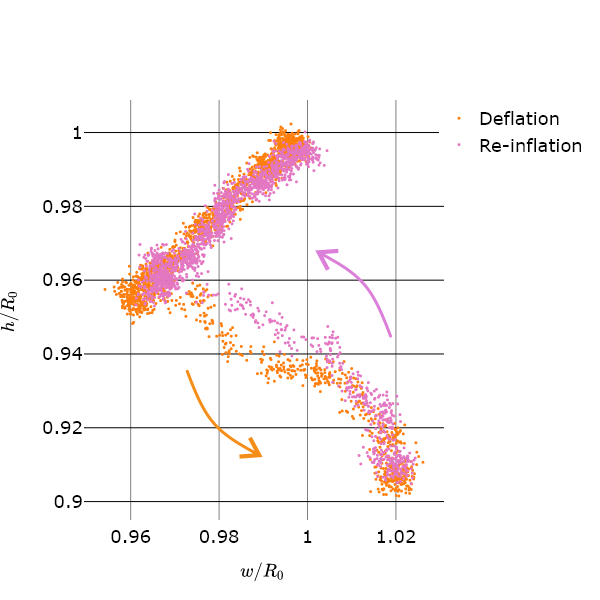}&  \includegraphics[width=0.5\textwidth]{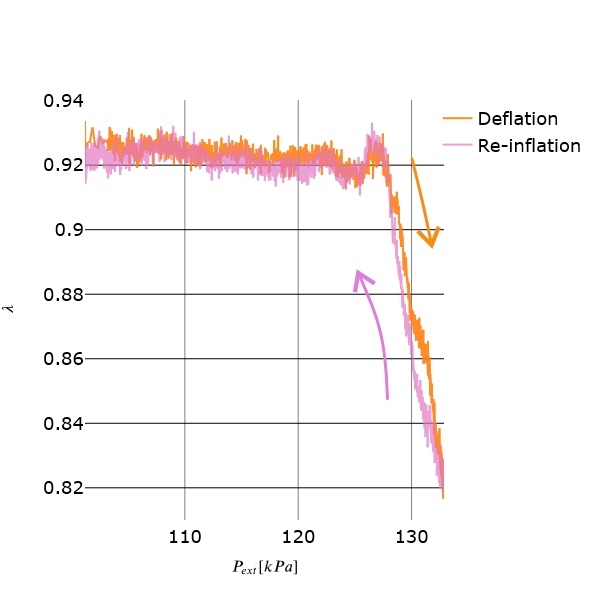} 
      \\
      (a)  & (b)  \\[6pt]
      \end{tabular}
\caption{Deformation of a $\chi_{dyn}=2$ N/m home-made shell along a slow external pressure cycle from atmospheric pressure $P_{\text{atm}}$ to $P_{\text{atm}}+\delta P$ : (a)  The  height $h$ as a function of the width $w$; both normalized with the initial radius $R_0$. (b)  Aspect ratio $\lambda$ as a function of applied pressure $P_{ext}$. Arrows indicate the time direction (increase then decrease of pressure).}
 \label{fig:typicaldef}
\end{figure*}

\begin{figure*}
\begin{tabular}{cc}
    \includegraphics[width=0.5\textwidth]{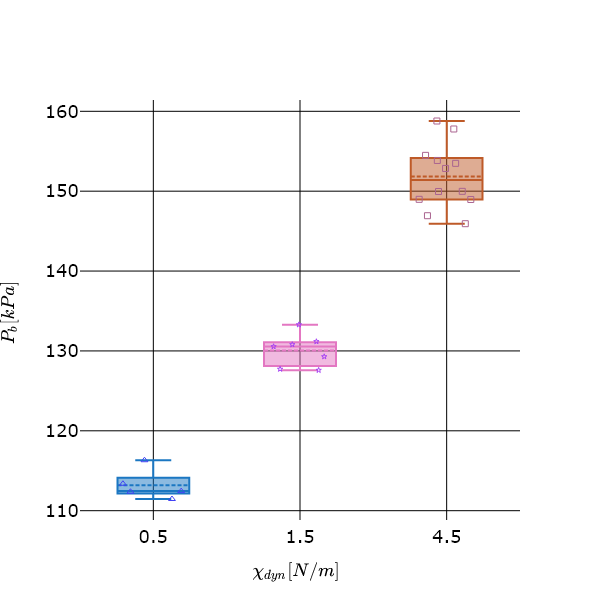}  &  \includegraphics[width=0.5\textwidth]{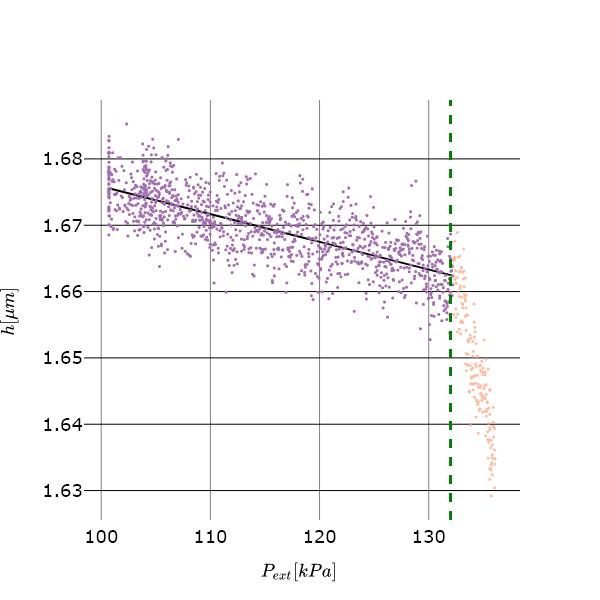}  
     \\
      (a)  & (b)  \\[6pt]

\end{tabular}
\caption{ (a) Buckling pressures $P_b$ for  three considered shells.  (b) Variation of the shell size $h$ with the applied pressure $P_{ext}$, for a shell with $\chi_{dyn}=1.5$ N/m. The spherical part (where $h$ is equal to the shell radius $R$, purple dots)  is linear and the corresponding fit is indicated in full line. The buckling $P_b$ corresponds to the change of slope, when the height $h$ of the shell decreases more abruptly (indicated with the dashed line). \label{fig:resPb}}\end{figure*}

Upon a slow increase (period of 40 s) of external pressure, a shell first shrinks isotropically (Fig. \ref{fig:typicaldef}(a)) then its aspect ratio suddenly drops for a given pressure threshold (Fig. \ref{fig:typicaldef}(b)), which is defined as the critical buckling pressure $P_{b}$. As shown in Fig. \ref{fig:resPb}(a), the buckling pressure increases with the shell dynamic stiffness. As for purely elastic shells \cite{djellouli2017}, an hysteresis in the $(h,w)$ space is visible; by contrast though, it is  not associated to a marked hysteresis in the $(\lambda,P_{ext})$ space.

The spherical deformation and the buckling process are now used to determine the mechanical properties of the shell, according to the model built in Sec. \ref{sec:model}.

According to Eq. \ref{eq:Pext-de-R_theo}, a linear fit $R=A+B P_{ext}$ of the  $R(P_{ext})$ curve in the spherical regime provides two equations between $P_{ref},R_{ref},\chi_{2D}$ and $\gamma$. An example of such a fit is shown in Fig. \ref{fig:resPb}(b).

Eq. \ref{eq:Rref} provides an equation for $R_{ref}$ in which the radius $R_b$ at buckling threshold is introduced, as well as the Poisson's ration $\nu$ and the characteristic length $d_{eff}$.  The buckling pressure $P_b$ is the pressure at which the aspect ratio drops (see Fig. \ref{fig:typicaldef}(b)), but this rough determination can be refined as the locus of the change of slope in the $h(P_{ext})$ curve, as shown in Fig. \ref{fig:resPb}(b). One ends up with 3 equations for 6 unknowns. In the following, we discuss general assumptions for three of them: $\nu$, $\gamma$ and $d_{eff}$.

\paragraph{Poisson's ratio:}  $\nu$ is, traditionally, difficult to measure with accuracy. We may nevertheless remark that, for usual materials, $\nu>0$: auxeticity, or negative Poisson's ratio, does not show up by chance in materials and are usually the result of sophisticated strategies. This bounds the prefactor $\sqrt{\frac{1-\nu}{3\left(1+\nu\right)}}$ to 0-1.7. For lipidic UCAs, Terzi et al. recently proposed a value $\nu=0.5$ \cite{terzi2019} ; however, this proposition emerges as the result of a calculation which is highly dependent on another parameter, the volume compressibility $K_{v}$, which is itself quite roughly estimated. Following a determination which resulted from the quantification of hundreds of shapes undertaken by comparable deformed thin spherical
shells of lipid monolayers in the gel phase \cite{quemeneur2012},
we will in this paper assume $\nu=0.8$, a value which in addition is compatible with the reputation of ``incompressibility'' (in the 2D sense) of lipid mono- or bilayers.
\paragraph{Surface tension:} the order of magnitude of such a value has to be at maximum the interfacial energy between a polar and an apolar liquid, {\itshape i. e.} $\approx 30$ mN/m. We remind that in surface tension based models, this surface tension is assumed to be close to 0, and we shall evaluate the impact of both these extreme values.
\paragraph{Characteristic length $d_{eff}$:}  in our experiments,  shells mostly present 3 folds after buckling (Fig. \ref{fig:shapes}). This sole observation discards the view of the shell leaving its spherical shape because of vanishing surface tension: in this case that would imply the absence of characteristic length, a wide zoology of multi-folded shapes would be observable.  According to \cite{quilliet12}, we can then deduce that $d_{eff}/R_{ref}=(\frac{0.940}{3})^2\simeq 10 \%$.
This is much larger than the thickness $d$ of the shell ($\approx\,5$ nm),
in a way comparable to what was observed on similar systems \cite{quemeneur2012} ; this difference reminds the importance of taking
into account the non isotropic nature of the material constituting
the shells. Indeed, lipid layers behave differently in the direction parallel
to the lipids, and in the two perpendicular directions, which are at
any place parallel to the shell ; such materials are called ``transverse isotropic'' \cite{lempriere68,chabouh2021}, and the shell's properties "orthoradially isotropic". In our analysis, we will check the impact of this value for $d_{eff}$, with respect to the customary choice  $d_{eff}=d$.\\

Table \ref{tab:databrut} synthesizes the parameters obtained from our analysis of  the raw data, and the mechanical parameters that are deduced from them, depending on the hypothesis that is made. Shells with similar sizes were selected, the dispersion in size is of order 0.1 microns. Mean values are reported in the Table. The quasi-linearity of the equations allows to  run our analysis on these data.

\begin{table}[!h]
\centering
\caption{Table of extracted parameters $A$ and $B$ from the spherical deflation of three types of home-made UCAs, of external buckling pressure $P_b$, and deduced buckling radius $R_b=A+B\,P_b$ and initial radius $R_0=A+ B\, P_{atm}$. From these data, shell properties $P_{ref}, R_{ref}$ and $\chi_{2D}$ are deduced, as well as the inner pressure $P_0$ in the initial state ($P_{ext}=P_{atm}$). Uncertainties on the fitting parameters $A$ (and radii) and $B$ are 2 nm and $0.01$ nm/kPa, respectively. $P_b$ is determined with an accuracy of 1 kPa.}
\label{tab:databrut}
\begin{tabular}{r|l|l|l|l}
\cline{3-5}
\multicolumn{2}{r|}{$\chi_{dyn}\rightarrow$}& { $0.55$ N/m} &{ $1.5$ N/m}&{$4.5$ N/m }\\ \cline{2-5}

\multirow{5}{*}{From raw data $\rightarrow$}&A ($\mu$m)&1.927   &1.655 &1.489 \\
&B (nm/kPa)&-1.304  &-0.462  &-0.372 \\
&$P_b$ (kPa)&113.07 &130.58&151.56 \\
&$R_b (\mu$m)&1.780 &1.595&1.433 \\

&$R_0 (\mu$m) &1.795&1.608 &1.451\\

\hline
Hypothesis $\downarrow$\\
\hline 
$d_{eff}/R_{ref}=10\%$ & $R_{ref} (\mu$m)&1.861 &1.643&1.476\\
 and $\gamma=30$ mN/m  &  $P_{ref}$(kPa) &82.4&63.1&76.3\\
 &  $P_{0}$ (kPa) &91.8&67.4&80.2\\
 & $ \chi_{2D}$ (N/m)&0.56&1.40&1.39\\
\hline
$d_{eff}/R_{ref}=10\%$ & $R_{ref}(\mu$m)&1.814&1.625&1.460\\
and $\gamma=0$ mN/m &  $P_{ref}$(kPa) &86.7&64.2&77.5\\
  &  $P_{0}$ (kPa) &89.5&66.2&78.9\\
 & $ \chi_{2D}$ (N/m)&0.51&1.35&1.35 \\
\hline
$d_{eff}=d=5$ nm & $R_{ref}(\mu$m)&1.833&1.614&1.450\\
and $\gamma=30$ mN/m   &  $P_{ref}$ (kPa) &105.2&125.1&145.0\\
&  $P_{0}$ (kPa) &112.0&126.5&144.7\\
 & $ \chi_{2D}$ (N/m)&0.51&1.27&1.27\\
\hline
$d_{eff}=d=5$ nm & $R_{ref}(\mu$m)&1.780&1.596&1.434\\
and $\gamma=0$ mN/m  &  $P_{ref}$ (kPa) &112.2&128.5&149.0\\
&  $P_{0}$ (kPa) &109.6&125.5&143.6\\
 & $ \chi_{2D}$ (N/m)&0.46&1.22&1.22\\
\end{tabular}
\vspace*{-4pt}
\end{table}


%
%
%

\section{Discussion}

The results obtained with our complete theoretical model are rich in lessons.

First, the different attempts summarized in Table \ref{tab:databrut} show that the determination of $\chi_{2D}$ is quite robust:  if $d_ {eff}$  and  $\gamma$ vary in the discussed range, $\chi_{2D}$ varies by at most 10 $\%$. This can be understood by looking at the relative weight of each term in the quasi-static elastic parameter $-B^{-1}$: the contribution from $\chi_{2D}$ is indeed dominant, hence it is directly related to the measurement of the slope in the $P_{ext}(R)$ curve.

This quasi-static elastic contribution is quite close to the dynamic elastic contribution  when the latter is equal to  $\chi_{dyn}=0.55$ N/m or $\chi_{dyn}=1.5$ N/m. However, for the $\chi_{dyn}=4.5$ N/m shell, the quasi-static elasticity does not match the dynamic one and is always three times smaller, whatever the hypothesis on the parameters is. Otherly stated, the quasi-static elasticity is quite similar for $\chi_{dyn}=1.5$ N/m and $\chi_{dyn}=4.5$ N/m shells.

On the other hand, buckling pressures show a clear increase with dynamic stiffness (Fig. \ref{fig:resPb}(a)). It is clear from the previous discussion that this increase of buckling pressure can not only be understood by an increase in stiffness $\chi_{2D}$.  Here, it is necessary to consider the value taken by the (reconstructed) inner pressure $P_0$ at initial radius $R_0$, whose determination strongly depends on the hypothesis made for the parameters $\gamma$ and, more importantly, $d_{eff}$.

The geometrical observation of 3-fold buckled shapes highlights the existence of a non-zero characteristic lengthscale $d_{eff}$, that we estimated to be in the order 10\% of the reference radius. The associated buckling pressure difference across the shell (Eq. \ref{eq:DeltaPc_BidiOnly}) implies that the reference pressure and more importantly the pressure $P_0$ in the initial state must be lower than atmospheric pressure (with some variations depending on the choice of surface tension). It also implies that at initial state, the shell is already under elastic compression ($R_{ref}>R_0$). This is not, a priori, incompatible with the mode of  production of  shells, which were first formed under subatmospheric pressure, and then lost part of their gas. Yet, this finding is in contrast with the usual output of surface tension based models that leads to the conclusion of initial surface tension in the order of 20-30 mN/m \cite{segers2018}, then to $P_0>P_{atm}$.  Finally, these results show that difference in the buckling pressure between the $\chi_{dyn}=1.5$ N/m and $\chi_{dyn}=4.5$ N/m shells does not lie in the difference in stiffness but rather in the difference in initial pressure, the later being larger for shells with higher dynamic stiffness.

Changing Poisson's ratio from 0.8 to 0.5 does not impact much the conclusions: the compression modulus is increased by some 5 $\%$ but it also implies a decrease of the initial pressure in the shell: for the $\chi_{dyn}=1.5$ N/m shell, in the first hypothesis, $P_0$ becomes 18 kPa, which may appear as a too strong asset. This observation tends to validate the choice of $\nu=0.8$ which was already suggested by a more extensive study on buckled shape of similar lipid coated shells \cite{quemeneur2012}.

{We finally remark that authors usually assume  that the inner pressure $P_0$ is equal to atmospheric pressure, when  spherical oscillations are analyzed to determine shell mechanical properties. This reduces the number of parameters to be calculated.} Here, if we impose this hypothesis and let $\gamma$ as an unknown parameter to be determined, we find no acceptable solution: for $\chi_{dyn}=1.5$ N/m, with 1st hypothesis ($d_{eff}/R_{ref}=10\%$): two solutions are found, $\gamma=-0.15 $ or 0.2 N/m, both being non physical. With the 3rd hypothesis ($d=5$ nm),  there is no solution for $\gamma$. As discussed in recent more advanced modeling papers for spherical oscillation, the values of initial pressure and of initial stress are key if one wishes to deduce correctly the other visco-elastic parameters of the shell \cite{chabouh2021,dash2022}. Buckling experiments may be a way to access these data.  

\section{Conclusion}

Our results highlight the difficulty in assessing a coherent view of the mechanical properties of lipidic microshells that would be relevant for all frequencies and amplitudes of loads, including triggering of buckling, or buckling-like, events.

Surface tension based models intrinsically assume  that the shell is always overpressurized, while buckling processes require, in the elastic shell framework, the inner pressure to be lower than the outer pressure. Here, we explained the buckling process under quasi-static load in the frame of the elastic theory, and showed that the overall elastic response in the spherical deformation under quasi-static load is similar to that obtained under pulsed, high amplitude, excitation, using a surface tension based model. However, we showed that coherence with known buckling thresholds and the typology of deformation with a small number of folds in our case, can only be explained by considering that vanishing surface tension is not the key to understand non-spherical deformations. Rather, an elastic contribution that includes an effective characteristic lengthscale that would emerge from the transverse anisotropy of the shell, is required to account for the observed phenomena. {The 3-fold typology of deformation is also observed in the widely used and characterized Sonovue shells, which leads us to anticipate that our conclusions would be valid for other types of lipid UCAs.}

The existence of the intermediate lengthscale also implies larger buckling thresholds than expected. This is similar to what was observed on millimetric armored bubbles  \cite{pitois2015,taccoen2016,Pitois2019}. Pitois et al. compared the threshold with what was expected from homogeneous material theory, with elastic moduli  measured independently on flat surfaces \cite{pitois2015}.  They found a buckling threshold 4-5 times higher than the theoretical threshold. In \cite{pitois2015,taccoen2016}, models are built to account for the observed buckling threshold, that takes into account the possibility for particles to rotate and do a kind of zig zag configuration, which allows to compress more the shells before buckling. This may inspire further modeling for buckling of lipid shells which present somehow a geometrically comparable configuration.  It would be interesting to understand if these local mechanisms could also explain the absence of hysteresis in  pressure-shape diagram,  which marks a strong difference with usual buckling-unbuckling paths for usual elastic shells. {A path towards a thorough modeling of these lipidic shells through both elastic and surface tension contributions may also have been opened by the recent work of Dash and Tamadapu, who introduced a curvature-dependent surface tension to describe the spherical oscillations of UCAs, allowing for an interesting analysis of existing experimental data for radial oscillations of UCAs.}

Finally, how quasi-static and dynamic experiments will be described by the same framework remains  an open question.  A hint may come from the study of Thomas et al., who observed an increase of the in-plane elasticity of lipid shells with the pressurization rate \cite{thomas2017}. Viscous phenomena must, in general, be taken into account \cite{hayman1981,Dykstra2022}. A macroscopic experiments showed recently that delay in buckling of visco-elastic shells is strongly related to the dynamics at the level of the defect where buckling initially nucleated \cite{Stein-Montalvo21}. Delay in buckling may lead to apparent perception of buckling threshold knock-up, by contrast with the usual knock-down observed in presence of defects \cite{lee16,hutchinson2016,Jimenez2017,yan2020,Paulose2013,Gerasimidis2020}.

In general, work is still needed to account for the buckling and/or rupture dynamics of lipid shells, which have recently shown an intensified non-linearity at high frequency and low excitation pressure amplitudes \cite{sojahrood2021}. In parallel, local speed of sound estimations in the buckled state have revealed an effective "softening" of the shells \cite{renaud2015,memoli2018}. While surface tension models propose to account for this softening by discarding the contribution of the shell (zero surface tension), a model that includes an elastic contribution from the shell can also reproduce this fact \cite{mokbel21}. Our present work encourages to dig into this direction.

\vskip6pt

\enlargethispage{20pt}

\ethics{Insert ethics statement here if applicable.}

\dataccess{Supporting data are available in a Dryad repository: https://doi.org/10.5061/dryad.bnzs7h4dx}

\aucontribute{BvE produced and characterized the home-made UCAs through ultrasounds. GCh designed the experimental protocole and set-up for buckling experiments. GCh and GCo carried out these experiments. GCh, CQ and GCo developed the buckling model. GCh, CQ and GCo performed the data analysis and interpretation. CQ and GCo conceived and designed the study. GCh, CQ, TS and GCo drafted the manuscript. All authors read, corrected and approved the manuscript.}
\competing{The author(s) declare that they have no competing interests.}

\funding{GCh, CQ and GCo acknowledge funding from CNRS through PEPS M\'ecanique du futur. GCo acknowledges funding from Labex Tec 21 for an outgoing grant. TS acknowledges financial support of the Max-Planck Center Twente.}

\ack{GCo acknowledges enlightening discussions with Douglas P. Holmes regarding the a priori non trivial effect of surface tension on buckling threshold. GCh would like to thank P. Marmottant for technical assistance. GCh appreciated the help of J. Fick in developing the tracking algorithm.}

\disclaimer{Insert disclaimer text here if applicable.}



\end{document}